\documentclass[]{mn2e}
\usepackage{graphicx}
\usepackage{psfig}
\usepackage{times}
\usepackage{amsmath}
\usepackage{amssymb}

\title[A New Simple Model for High Frequency QPOs in Black Hole
Candidates]
{A New Simple Model for High Frequency Quasi Periodic Oscillations
in Black Hole Candidates}

\author[Rezzolla, Yoshida, Maccarone and Zanotti]
	{L. Rezzolla$^{1,2}$, S'i. Yoshida$^{1,2}$ 
	T. J. Maccarone$^{1,3}$ and O. Zanotti$^{1}$		\\
								\\
	$^1$SISSA, International School for Advanced Studies,	
        Via Beirut, 2 34014 Trieste, Italy			\\
	$^2$INFN, Sezione di Trieste, Via Valerio, 2 34127 	
	Trieste, Italy						\\
	$^3$Astronomical Institute ``Anton Pannekoek'',
	University of Amsterdam, Kruislaan 403, 1098 SJ, Amsterdam,
	The Netherlands
	}

\begin{document}

\maketitle
\pagerange{\pageref{firstpage}--\pageref{lastpage}}\pubyear{2003}

\label{firstpage}

\begin{abstract}
	Observations of X-ray emissions from binary systems have long
	since been considered important tools to test General Relativity
	in strong-field regimes. The high frequency quasi-periodic
	oscillations (HFQPOs) observed in binaries containing a black hole
	candidate, in particular, have been proposed as a means to
	measure more directly the black hole properties such as its mass
	and spin. Numerous models have been suggested to explain the HFQPOs
	and the rich phenomenology accompanying them. Many of these
	models rest on a number of assumptions and are at times in
	conflict with the most recent observations. We here propose a
	new, simple model in which the HFQPOs result from basic $p$-mode
	oscillations of a small accretion torus orbiting close to the
	black hole. We show that within this model the key properties of
	the HFQPOs can be explained simply, given a single reasonable
	assumption. We also discuss observational tests that can falsify
	the model.
\end{abstract}


\date{Accepted 0000 00 00.
      Received 0000 00 00.}

%-------------------------------------------------------------
\section{Introduction}
\label{intro}
%-------------------------------------------------------------

	One of the strongest motivations for studying X-ray binaries has
been the hope that these systems could be used as probes of fundamental
physics.  Stellar mass black holes represent a laboratory for studying
strong field General Relativity, while neutron stars allow us to test
equations of state for nuclear matter. The high frequency quasi-periodic
oscillations (HFQPOs) seen from these sources have been held forth as one
of the most promising diagnostics, with the potential to measure, for
example, the masses and radii of neutron stars (e.g. Kluzniak et al.,
1990; Miller et al., 1998) or the masses and spins of black holes
(e.g. Wagoner et al. 2001; Abramowicz and Kluzniak 2001).

	A great deal more work has been done explaining the HFQPO
phenomena in neutron star systems than those in black hole systems, and
this is motivated in large part by the much larger number of observed
systems and the richer data sets of these systems.  A correlation between
the frequencies of the high frequency QPOs and the break frequency for
the broadband noise component of the Fourier power spectrum that fits
both neutron stars and black holes (Psaltis, Belloni, van der Klis 1999 -
PBK) has been suggested to provide evidence that the same mechanism must
be taking place in all these systems.  The extension of this correlation
to include white dwarf systems (Mauche 2002) bolstered these claims.

	If the PBK and Mauche (2002) correlations are indeed universal
and the interpretation that a single mechanism is at work in all cases is
correct, then the mechanism cannot depend on the presence of a stellar
surface, of a strong magnetic field, or of a strong (i.e. relativistic)
gravitational field.  Instead, a model taking advantage of the properties
of a non-uniform rotating fluid would be a strong candidate for
explaining the HFQPOs (see e.g. Titarchuk 2003; Osherovich and Titarchuk
1999).

More recently, though, additional phenomenology has emerged which
indicates some fundamental differences between the neutron star and black
hole systems, and may suggest that different models apply to these
sources after all.  The kilohertz QPOs are typically seen in multiples.
In the neutron star systems, the separation in the HFQPO frequencies is,
in general, nearly constant, with the frequency separation shrinking as
the frequencies increase (see e.g. van der Klis et al., 1997; Mendez et
al., 1998). In the black hole systems, on the other hand, the kilohertz
QPO frequencies seem to drift by much smaller amounts (Strohmayer 2001a,
2001b) and to be found in ratios of small integers (i.e. 1:2, 2:3, or
1:2:3; Abramowicz and Kluzniak, 2001; Remillard et al., 2002; Homan et
al. 2003b). There are also some claims for harmonic structure in XTE J
1650-500 (Homan et al. 2003a), with peaks seen at 110, 140, 210, and 270
Hz, but the identification of a harmonic structure is not as clear here
as the frequencies are not all identified simultaneously and, in fact,
seem to drift. In addition to this, recent observations of the probable
black hole transient XTE J 1550-564 indicate that its HFQPOs do not
always fit on the PBK correlation (Remillard et al., 2002). These
differences suggest that the PBK correlation may not apply to all of the
HFQPOs seen; models requiring a stellar surface or general relativistic
effects, or both (e.g. Stella and Vietri 1998) need not be rejected.  As
we wish to explain the integer ratios of the frequencies of HFQPOs from
the dynamically confirmed black hole candidates, we will concentrate here
on models which do not require a stellar surface.

	A model applied primarily to the HFQPOs from black hole
candidates is the ``discoseismic'' model which asserts that $g$ modes
should become trapped in the potential well of a Keplerian disc in a Kerr
potential (e.g. Nowak et al., 1997). The size of the region where the
modes are trapped depends on both the mass and the spin of the accreting
black hole.  Additional frequencies of oscillation should be expected
from $p$ (pressure) modes and $c$ (corrugation) modes. The predictions of
the model are well summarised by Kato (2001). Given pairs of high
frequency QPOs and a proper identification of the frequencies with the
particular modes, one can measure both the black hole mass and spin to
relatively high accuracy (Wagoner et al., 2001 and references
therein). The discoveries of three systems where the HFQPOs show a
harmonic structure with relatively strong peaks seen in integer ratios
1:2, 2:3, or 1:2:3 seem to cast some doubt upon this model. However, the
discoseismic model remains viable for the intermediate frequency QPOs in
GRS 1915+105, seen at 67 Hz (Morgan et al., 1997) and 40 Hz (Strohmayer,
2001b). We emphasize that such harmonic structure has been seen only from
systems thought to contain a black hole as the compact accretor.

	Being the first to point out that the frequencies of QPOs in some
black hole and neutron star sources were in a ratio of small integers,
Abramowicz and Kluzniak (2001, 2002, 2003a) have proposed the
``resonance'' model, in which a harmonic relationship in the HFQPOs
frequencies can be produced as a result of orbital resonances. In
particular, the model suggests that an initial perturbation is amplified
at a radius where the radial epicyclic frequency for point-like masses is
in resonance with the latitudinal epicyclic frequency, with the two
frequencies being in (small) integer ratios (In a Schwarzschild spacetime
the latitudinal epicyclic frequency and the orbital one coincide.). These
annuli tend to be close to the black hole event horizon for the observed
frequencies and, hence, given a mass estimate for the black hole and an
identification of the ratio of the frequencies in resonance, a black hole
spin could be measured.

	It should be noted that given the observed frequencies of 162 and
324 Hz in GRS 1915+105 it is not possible to produce the 40 and 67 Hz
QPOs with a discoseismic model and the higher frequency QPOs with a
resonance model while retaining the same values for the black hole mass
and spin (Maccarone 2002). Another potential problem for these two last
models is the observational evidence for a frequency ``jitter'' in the
HFQPOs of XTE~J1550-564, i.e. for small variations of about 10\% in
frequencies for about 15\% of the time. This difficulty could be
particularly severe for the resonance model, whose relevant frequencies
confine the resonance to a narrow region in radial coordinates [This
situation is worsened for the 1:3 resonance for which the radial
variation of the frequencies is even more rapid (Remillard et
al., 2002).]. More recently, however, Abramowicz et al., (2003b) have
considered a perturbative approach to the standard resonance model and
were able to show that the resonance can take place also near the radial
positions at which the epicyclic frequencies are in an exact 2:3
ratio. While this result offers at least in part a possible explanation
for the frequency jitter, the radial extension over which the resonance
takes place (which is $\sim 0.1\, GM/c^2$ at most, Karas 2003) may still
be too small to produce the observed modulation in the emissivity.

	We here propose an explanation for the high frequency QPOs in
black hole candidates that has connections with both the discoseismic and
the resonance model. More precisely, as in the discoseismic model, we are
here focussed on global oscillation modes of a fluid orbiting in the
vicinity of the black hole and, as the resonance model, we consider
fundamental the observational evidence of a 2:3 ratio in the HFQPO
frequencies. Our model, however, has distinct novel features and, most
importantly, is based on the existence of a non-Keplerian (geometrically
thick) disc orbiting in the vicinity of the rotating black hole. Using
this {\it single} assumption, we show that the HFQPO phenomenology finds
a simple explanation.

	In Section 2 we introduce the basic properties of our model and
discuss how, using this single assumption, we can explain the most
important aspects of the HFQPO phenomenology. Finally, in Section 3, we
explain how the model can be used to deduce the black hole properties and
the ways in which it can be refined or refuted.

%-------------------------------------------------------------
\section{The Model}
\label{nr_and_tm}
%-------------------------------------------------------------

	In contrast to a Keplerian accretion disc, which is in principle
infinitely extended, a non-Keplerian disc can easily be constructed to
have a finite size, the extent being determined uniquely by the
distribution of the specific angular momentum and by the pressure
gradients. Because pressure gradients play such an important role,
non-Keplerian discs tend to be geometrically thick and look like tori
rather than thin discs. Depending then on the pressure gradients and
angular momentum distribution, the orbiting fluid is confined to a
finite-size region which can behave as a cavity in which global
oscillation modes could be trapped. This is a fundamental difference from
Keplerian discs, which have no definite outer boundary and in which
outward propagating waves cannot grow (Kato 2001).

	When a disc (either Keplerian or non) initially in equilibrium is
perturbed in some way, restoring forces appear to compensate for the
perturbation. A first restoring force is the centrifugal force, which is
responsible for inertial oscillations of the orbital motion of the disc
and hence for epicyclic oscillations (this is the restoring force at work
in the resonance model). A second restoring force is the gravitational
field in the direction vertical to the orbital plane and which will
produce a harmonic oscillation across the equatorial plane if a portion
of the disc is perturbed in the vertical direction (these oscillations
are extensively studied in galactic dynamics). A third restoring force is
provided by pressure gradients and the oscillations produced in this way
are closely related to the sound waves propagating in a compressible
fluid. In a geometrically thick disc, the vertical and horizontal
oscillations are in general coupled and more than a single restoring
force can intervene for the same mode. A detailed discussion of how to
classify the different modes of oscillation through its main restoring
force can be found in Kato (2001), but it is here sufficient to remind
that $c$ modes are essentially controlled by the vertical gravitational
field, that $g$ modes are mainly regulated by centrifugal and
pressure-gradient forces, and that all of the restoring forces discussed
above play a role in the case of $p$ modes. It is also useful to
underline that because we are here interested in modes with a prevalent
horizontal motion and frequencies above the epicyclic one, we are
essentially selecting ``inertial-acoustic'' modes having centrifugal and
pressure gradients as only restoring forces. Hereafter we will refer to
these simply as $p$ modes.

	Following a recent investigation of the nonlinear dynamics of
perturbed relativistic tori orbiting around a Schwarzschild black hole
(Zanotti et al., 2003), we have analysed the {\it global} oscillation
properties of such systems and, more specifically, we have performed a
perturbative analysis of the axisymmetric modes of oscillation of
relativistic tori in a Schwarzschild spacetime and in the Cowling
approximation (Rezzolla et al., 2003a). The eigenvalue problem that needs
to be solved to investigate consistently the oscillation properties of
fluid tori is simplified considerably if the vertical structure is
accounted for by an integration in the vertical direction. Doing so
removes one spatial dimension from the problem, which can then be solved
integrating simple ordinary differential equations. While an
approximation, this simpler model reproduces with a precision of less
than a few percent, the numerical results obtained with fully nonlinear,
2D numerical calculations (see Rezzolla et al., 2003a for a comparison).

	One of the important results of the perturbative investigation
was that the eigenfunctions and eigenfrequencies found were those
corresponding to the $p$ modes of the torus. Furthermore, the behaviour
of the fundamental frequencies was observed to converge to the epicyclic
frequency $\kappa_{\rm r}$ at the position of maximum density in the
torus $r_{\rm max}$ as the torus's size was progressively reduced to zero
(see Fig. 4 of Rezzolla et al., 2003a). Finally and most importantly, the
eigenfrequencies computed both for the fundamental mode of oscillation as
well as for the first few overtones, were found to be in a harmonic
sequence 2:3:4:$\ldots$ to within 5--10\%, the exact value depending on
the specific model for the torus (in general larger tori have
progressively smaller ratios). Note that such a relation among the
eigenfrequencies is not a standard property of $p$-mode oscillations. In
stars, for instance, this happens only for purely radial
oscillations. Despite showing this important property, the fundamental
frequencies were smaller than the observed lower HFQPO frequencies when
the estimated masses of the black hole candidates were used.

	Stimulated by this mismatch, we have extended the calculations to
the case of a torus orbiting around a Kerr black hole. The details on
these calculations will be presented in a separate paper (Rezzolla, and
Yoshida, 2003b) but the mathematical setup and the numerical methods used
follow closely those presented in Rezzolla et al. (2003a). Here, we will
concentrate on summarizing the results and discussing how they can be
used to construct a new model to explain the HFQPOs in black hole
candidates.

\begin{figure}
\centerline{ \psfig{file=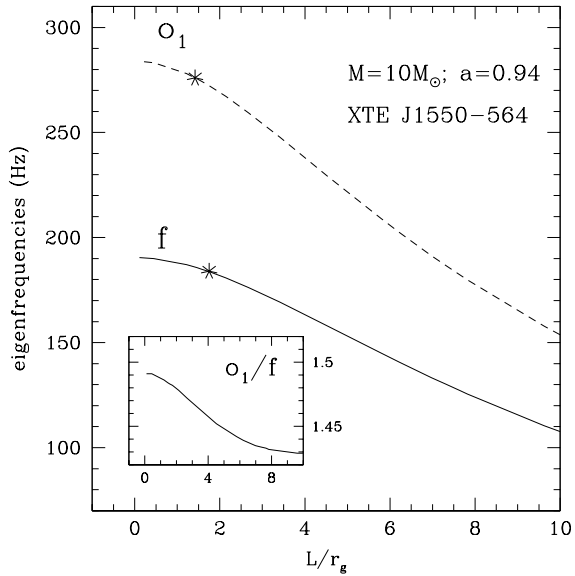,angle=0,width=8.cm} }
\caption{
\label{fig1}
$p$-mode frequencies for a non-Keplerian disc orbiting a Kerr black hole
with mass $M=10\ M_\odot$ and spin $a=0.94$. The fundamental frequencies
$f$ are indicated with a solid line, while the first overtones $o_1$ with
a dashed line. The two asterisks show the values of the HFQPOs observed
in XTE~J1550-546, while the inset shows the ratio of the two
frequencies.}
\end{figure}

	Figure~\ref{fig1} shows a typical example of the solution of the
eigenvalue problem for a black hole with mass $M=10\ M_{\odot}$ and its
comparison with the observations of XTE J1550-564 (other sources could
equally have been used). In particular, we have plotted the value of the
different eigenfrequencies found versus the radial extension $L$ of the
torus, expressed in units of the gravitational radius $r_g\equiv
GM/c^2$. The sequences have been calculated for a distribution of
specific angular momentum following a power-law (see Rezzolla et al.,
2003a for details), keeping constant $r_{\rm max}=3.489$, and for a black
hole with dimensionless spin parameter $a \equiv J/M^2=0.94$ to maximize
the value for $L$. Indicated with a solid line are the fundamental
frequencies $f$, while the first overtones $o_1$ are shown with a dashed
line; each point on the two lines represents the numerical solution of
the eigenvalue problem. The asterisks represent the frequencies of the
HFQPOs detected in XTE~J1550-564 at 184 and 276 Hz, respectively. Note
that the two plotted eigenfrequencies are close to a 2:3 ratio over the
full range of $L$ considered (this is shown in the inset) and that while
they depend also on other parameters in the problem (e.g. the position of
$r_{\rm max}$, the angular momentum distribution, the polytropic index),
these dependences are very weak so that the frequencies depend
effectively on $M$, $a$ and $L$. As a result, if $M$ and $a$ are known, a
diagram as the one shown in Fig.~\ref{fig1} could be used to determine
the dimension of the oscillating region $L$ rather accurately.

	On the basis of these results, we suggest that the HFQPOs observed
in black hole candidate systems can be interpreted in terms of {\it
$p$-mode oscillations of a small-size torus orbiting in the vicinity of a
black hole}. Such a configuration could be produced whenever an
intervening process (e.g. large viscosity, turbulence, hydrodynamical and
magnetohydrodynamical instabilities, etc.) modifies the Keplerian
character of the flow near the black hole. 

	Two remarks are worth making at this point. The first one is
about the existence of a non-Keplerian fluid motion: this is required
{\it only very close} to the black hole (the inner edge of the torus can
be in principle be located at the marginally bound orbit) and beyond this
region the fluid motion can be Keplerian. Stated differently, the HFQPOs
observed could be produced by the inner parts of a standard,
nearly-Keplerian, geometrically thin accretion disc, where a variety of
physical phenomena can introduce pressure-gradients. The second remark is
about the stability of these tori to non-axisymmetric oscillations. It is
well-known that a stationary (i.e. non-accreting) perfect fluid torus
flowing in circular orbits around a black hole is subject to a dynamical
instability triggered by non-axisymmetric perturbations (Papaloizou and
Pringle, 1984). It is less well-known, however, that the instability can
be suppressed if the flow is non-stationary. Stated differently, a fluid
torus around black could be stable to non-axisymmetric perturbations if
mass-accretion takes place (Blaes 1987; Blaes and Hawley 1988; Hawley
1991; De Villiers and Hawley 2002). Because the tori discussed here are
assumed to be the terminal part of standard accretion discs, we expect
them to be stable to non-axisymmetric perturbations as long as magnetic
fields are unimportant.

	In what follows we discuss how the HFQPO phenomenology finds simple
explanations within this model.

\begin{description}

	\item {\it (i)} The harmonic relations between the HFQPO
	frequencies in black hole candidates are {\it naturally}
	explained within this model. When the torus is sufficiently
	small, in fact, it can be thought of as a cavity in which the $p$
	modes effectively behave as trapped sound waves. If the sound
	speed in the cavity were constant, the frequencies of these
	standing waves would be in an exact integer ratio. In reality the
	sound speed is not constant but the eigenfrequencies found are in
	a sequence very close to 2:3:4 in the parameter range we are
	interested in.

	\item{\it (ii)} Being global modes of oscillation, the same
	harmonicity is present at {\it all} radii within the torus. This
	removes the difficulty encountered in the resonance model and
	provides also a larger extent in radial coordinates where the
	emissivity can be modulated (cf. Fig.~\ref{fig1}).

	\item {\it (iii)} By construction, the frequencies scale like
	$1/M$. This is in agreement with the observations made of
	XTE~J1550-564 and GRO~J1655-40 as long as the spins are similar
	(Remillard et al., 2002). On the other hand, a rather narrow
	range of black hole spins has been suggested as a possible
	explanation for the narrow range of radio-to-X-ray flux ratios in
	the Galactic X-ray binary systems (Fender 2001).

	\item {\it (iv)} The frequency jitter can be naturally
	interpreted in terms of variations of the size of the oscillating
	cavity $L$. Indeed, the frequencies may drift {\it arbitrarily}
	over a large range with the harmonic structure preserved
	(cf. Fig.\ref{fig1}).

	\item {\it (v)} The observed variations in the relative strength
	of the peaks can be explained as a variation in the perturbations
	the torus is experiencing. (This has been reproduced with
	numerical simulations; Rezzolla et al., 2003a). Furthermore, while
	the low frequency overtones are energetically favoured and the
	corresponding eigenfunctions possess less nodes, {\it any number}
	of harmonics could in principle be observed.

	\item {\it (vi)} The evidence that an overtone can be stronger
	than the fundamental in the harder X-ray bands (Strohmayer
	2001a), can be explained simply given that the overtone is an
	oscillation preferentially of the innermost (and hottest) part of
	the accretion flow (see Rezzolla et al., 2003a for the
	eigenfunctions).

\end{description}

%-------------------------------------------------------------
\section{Black hole properties and model tests}
\label{ot}
%-------------------------------------------------------------

	Because the $p$ modes discussed here represent the basic
oscillation modes of the torus, some of their features will not be very
sensitive to general relativistic corrections. The harmonic relation
between the fundamental frequency and the first overtones is one of such
features and this has been encountered not only in Schwarzschild and Kerr
spacetimes, but also when we have considered a simple Newtonian
gravitational potential (Rezzolla et al., 2003a). While these {\it
qualitative} features remain unchanged in relativistic regimes, {\it
quantitative} difference emerge and since these depend on the mass and
spin of the black hole, they can be used to measure the black hole
properties. This weak dependence on relativistic effects is an important
way in which our model for HFQPOs differs from the ones proposed so far.
	
\begin{figure}
\centerline{ \psfig{file=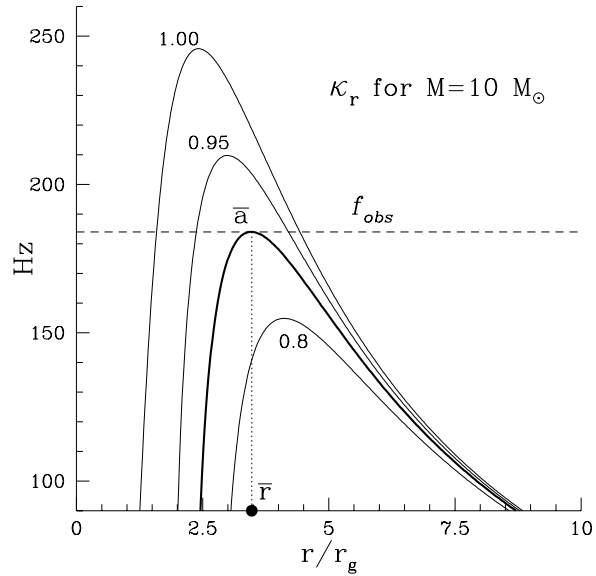,angle=0,width=8.cm} }
\caption{
\label{fig2}
Schematic diagram showing how to derive a lower limit on the black hole
spin ${\bar a}$, and an upper limit on the torus size ${\bar r}$, by
comparing the lower HFQPO frequency with the radial epicyclic frequency
$\kappa_{\rm r}$. Different curves show $\kappa_{\rm r}$ for different
values of $a$.}
\end{figure}

	As discussed above, the solution of the eigenvalue problem in a
Schwarzschild spacetime has shown that the fundamental $p$-mode frequency
tends to the radial epicyclic frequency at $r_{\rm max}$ in the limit of
a vanishing torus size. For a Kerr black hole, this frequency is a
function of its mass and spin through the relation {$\kappa^2_{\rm
r}(a,M) =(GM/r^3)(1+a/{\hat r}^{3/2})^{-2}(1 - 6/{\hat r} + 8a/{\hat
r}^{3/2}- 3a^2/{\hat r}^2)$}, where ${\hat r}\equiv r/r_g$ (Kato
2001). Exploiting this property and the solution of the eigenvalue
problem for a large number of different models for the tori, (i.e. with
different pressure and density profiles, distributions of specific
angular momentum, polytropic indices, etc.) we suggest that a first
estimate of the black hole spin and of the size of the oscillating torus
can be obtained once the lower frequency in the HFQPOs is measured
accurately. In this case, and as shown schematically in Fig.~\ref{fig2},
given a black hole candidate with measured mass $M_*$, a {\it lower
limit} on the black hole spin ${\bar a}$ can be deduced as the value of
$a$ at which the maximum epicyclic frequency is equal to the lower
observed HFQPO frequency, i.e.  ${\bar a} \lesssim a \leq 1$, with
max$[\kappa_{\rm r}({\bar a},M_*)]=\;$lower HFQPO frequency. Because for
small tori $L \lesssim r_{\rm max}$ (cf. Rezzolla et al., 2003a), once
${\bar a}$ has been determined, the radial position ${\bar r}$ of the
maximum of the epicyclic frequency provides an {\it upper limit} on the
size of the torus, i.e. $0 \leq L \lesssim {\bar r}$, where $d\kappa_{\rm
r}({\bar a},M_*)/dr = 0$ at $r={\bar r}$. These two estimates should be
considered first approximations only and should be used to further define
the properties of the oscillating region; in the case of XTE~J1550-564,
for instance, the values estimated in this way are ${\bar r} \simeq
3.49\, r_g$, and $a\simeq 0.89$ (cf. Table 1), while the solution of the
full eigenvalue problem built around these estimates yields the more
accurate limits: ${\bar r} \simeq 1.75\, r_g$, and $a\simeq 0.94$
(cf. Fig.~\ref{fig1}).  The values of ${\bar a}$ and ${\bar r}$ for the
black hole candidates with HFQPO observations and for which an estimate
of the mass is available, are presented in Table 1, which also shows a
surprisingly small scatter. Given these estimated sizes and the location
of the inner and outer edges of the tori, it is easy to calculate that
the gravitational potential energy change across the torus is $\gtrsim
6\%$ of the energy at the outer edge; this relative energy loss could
account for the modulated emission observed in QPOs.

	We note that this model requires rather large values of the spin
parameter. This is not too surprising given that the massive star
progenitors of the black holes in X-ray binaries routinely have more
angular momentum than even a maximally rotating black hole of the mass of
the remnant (see, for instance, the velocity measurements in Brown,
Vershueren 1997).  The large required spins also make it unlikely that
this model will be successfully applicable to neutron star HFQPOs; we
re-emphasize that this does not represent a major problem for this model,
since the neutron star systems do not show simple integer ratios of
frequencies as part of the phenomenology of their HFQPOs.

\begin{table}
\label{tab2}
\begin{center}
\begin{tabular}{lcccc}\hline
Source & $M/M_{\odot}$ & $f$ (Hz) & ${\bar a}$ 
       & ${\bar r}/r_g$  			\\
 \hline 					\\
GRS~1915+105  & 14.0 & 162 & 0.9748 & 2.739 	\\
XTE~J1550-564 & 10.0 & 184 & 0.8909 & 3.489 	\\
GRO~J1655-40  & ~6.3 & 300 & 0.9035 & 3.391  	\\
 ~ 						\\
\hline 
\end{tabular}
\caption{ \small{Lower limits on the black hole spin ${\bar a}$, and
upper limits on the torus size ${\bar r}$, as estimated from the lower
HFQPO frequency of black hole candidates.}}
\label{freq-lin}
\end{center} 
\end{table}

	It is also worth noting that if this model is confirmed to be
correct, it will give us a reliable means of placing an upper limit on
the mass of a black hole while at the same time requiring the existence
of a black hole (see Abramowicz et al., 2002).  Observing relatively
high-frequency QPOs in pairs with small integer ratios would give us good
suggestive evidence for the existence of low-mass black holes.  At the
present, in fact, there is no strong evidence for any compact objects
with masses between the $\simeq 1.4\; M_\odot$ of a neutron star and the
$\simeq 6.3\; M_\odot$ of GRO J 1655-40.

	A final comment will be reserved to the possibility of finding
support to this model from the observational data. There are now
extensive observations of these black hole HFQPOs and interesting
spectral and flux correlations are beginning to be found (see, for
instance, Remillard et al., 2002a, and references therein). These
observations could be used to deduce the presence of a small torus in the
terminal part of the accretion discs. Furthermore, being based on a
single assumption, our model for HFQPOs is simply constructed, but,
equally simply, it can be confuted. We now discuss two observational
constraints that if not met will cast serious doubts about whether this
model is a satisfactory description of the phenomenology observed in
HFQPOs. The first constraint can be easily deduced from Fig.~\ref{fig2}
and basically requires the lower HFQPO frequency to be always less than
the maximum possible epicyclic frequency for a black hole of mass $M_*$,
i.e. lower HFQPO frequency$\;\leq{\rm max}[\kappa_{\rm r}(a =
1,M_*)]$. The second constraint, on the other hand, is that even if the
HFQPOs frequencies change as a result of the change in size of the torus,
they should nevertheless appear in a harmonic ratio to within
5-10\%. Both of these observational requirements are sufficiently
straightforward to assess and therefore provide a simple and effective
way of falsifying this model.

\section*{Acknowledgments}

It is a pleasure to thank L. Stella, O. Blaes, M. Klein-Wolt and
M. Mendez for helpful comments, M.  Abramowicz, W. Kluzniak, and V. Karas
for a comparison with the resonance model. Financial support for this
research has been provided by the MIUR and by the EU Network Programme
(Research Training Network Contract HPRN-CT-2000-00137). LR also
acknowledges hospitality at the KITP in Santa Barbara where this work was
completed (NSF grant PHY99-07949).

\label{lastpage}  
\end{document}